\documentstyle[12pt,epsf]{article}
\begin{document}
\def\lsim{\mathrel{\lower4pt\hbox{$\sim$}}\hskip-12pt\raise1.6pt\hbox{$<$}\;}
\def\BAR{\bar}
\def\fm{{\cal M}}
\def\fl{{\cal L}}
\def\n{\delta}
\def\gsim{\mathrel{\lower4pt\hbox{$\sim$}}
\hskip-10pt\raise1.6pt\hbox{$>$}\;}

\vspace*{-.5in}
\rightline{BNL-HET-99/26}
\begin{center}

{\large\bf 
Graviton Production By 
Two Photon  and Electron-Photon Processes 
In Kaluza-Klein Theories With Large Extra Dimensions\\
}

\vspace{.3in}

David Atwood$^{1}$\\ 
\noindent Department of Physics and Astronomy, Iowa State University, Ames,
IA\ \ \hspace*{6pt}50011\\
\medskip

Shaouly Bar-Shalom$^{2}$\\ 
\noindent Department of Physics, University of California, Riverside, 
CA 92521\\
\medskip
and\\
\medskip

Amarjit Soni$^{3}$\\
\noindent Theory Group, Brookhaven National Laboratory, Upton, NY\ \ 
11973\\
\footnotetext[1]{email: atwood@iastate.edu}
\footnotetext[2]{email: shaouly@phyun0.ucr.edu}
\footnotetext[3]{email: soni@bnl.gov}
\end{center}
\vspace{.2in}

\begin{quote}
{\bf Abstract:}

We consider the production of gravitons via two photon and electron-photon
fusion in Kaluza-Klein theories which allow TeV scale gravitational
interactions.  We show that at electron-positron colliders, the processes
$\ell^+\ell^-\to \ell^+\ell^-+graviton$, with $\ell=e$, $\mu$, can lead to
a new signal of low energy gravity of the form $\ell^+\ell^-\to
\ell^+\ell^-+missing~ energy$ which is well above the Standard Model
background.  For example, with two extra dimensions at the Next Linear
Collider with a center of mass energy of 500 or 1000 GeV, hundreds to
thousands such $\ell^+\ell^-+graviton$ events may be produced if the scale
of the gravitational interactions, $M_D$, is around a few TeV. At a
gamma-electron collider, more stringent bounds may be placed on $M_D$ via
the related reaction $e^-\gamma\to e^-G$. For instance, if a $1~TeV$
$e^+e^-$ collider is converted to an electron-photon collider, a bound of
$\sim 10~TeV$ may be placed on the scale $M_D$ if the number of extra dimensions 
$\delta=2$ while a bound of $\sim 4~TeV$ may be placed if $\delta=4$.

\end{quote} 
\vspace{.3in} 
\newpage

\section{Introduction}

Gravity is the weakest force of nature and,
although it ultimately controls the shape of the entire universe, its role
in fundamental interactions remains obscure. This is due to the fact that
gravity remains weak until the unreachably high scale of the Planck mass
and thus there is no experimental data to construct a theory of gravity at
small distances.

Of course the lack of experimental evidence has not deterred the
construction of theories to account for the properties of gravitation at
short distances. In this Letter we will consider certain Kaluza-Klein
theories which contain additional compact dimensions besides the
four space-time dimensions.

In such theories it was traditionally assumed that the compact dimensions
form a manifold which is unobservably small (perhaps at the Planck scale)
and thus remain hidden. However, recent advances in
M-theory~\cite{mtheory}, a Kaluza-Klein theory in which there are 11 total
dimensions, suggest another possible scenario~\cite{add1}.  In models
proposed in~\cite{add1,add2}, $\delta$ of these extra dimensions may be
relatively large while the remaining dimensions are small. In this class
of theories, the known fermions, the strong, weak and electromagnetic
forces exist on a 3-brane while gravity may act in $4+\delta$ dimensions.
The size of these extra dimensions, $R$, is related to an effective Planck
mass, $M_D$, according to~\cite{add1}:

\begin{eqnarray}
8\pi R^\delta M_D^{2+\delta}\sim M_P^2
\label{Rsize}
\end{eqnarray}

\noindent
where $M_P=1/\sqrt{G_N}$ is the Planck mass and
$G_N$ is Newton's constant.
Indeed the effective Planck mass at which gravitational effects become
important may be as small as $O(1~{\rm TeV})$ in which case such effects
may be probed in collider experiments.

In this scenario, at distances $d<R$ the Newtonian inverse square law will
fail~\cite{add1}. If $\delta=1$ and $M_D=1$ TeV, then R is of the order
of $10^{8}~km$, large on the scale of the solar system, which is clearly
ruled out by astronomical observations.  However, if $\delta\geq 2$ then
$R< 1~mm$; there are no experimental constraints on the behavior of
gravitation at such scales~\cite{cavin} so these models are possible.

Astonishingly enough if $M_D\sim 1$ TeV then gravitons may be readily
produced in accelerator experiments.  This is because the extra dimensions
give an increased phase space for graviton radiation. Another way of
looking at this situation is to interpret gravitons which move parallel to
the 4 dimensions of space time as the usual gravitons giving rise to
Newtonian gravity while the gravitons with momentum components
perpendicular to the brane are effectively a continuum of massive objects.
The density of gravitons states is given by~\cite{add1,add2,wells,taohan}
where in particular we use the convention of~\cite{wells,convention}:

\begin{eqnarray}
D(m^2)={dN\over d m^2}={1\over 2} S_{\delta-1} 
{\BAR M_P^2 m^{\delta-2}\over M_D^{\delta+2}}
\label{density_of_states}
\end{eqnarray}

\noindent
where $m$ is the mass of the graviton, $\BAR M_P=M_P/\sqrt{8\pi}$
and 
$S_{\delta-1}=2\pi^{\delta/2}/\Gamma[\delta/2]$.
The probability of graviton emission  may thus become large when the sum
over the huge number of graviton modes is considered. 

Gravitons with polarizations that lie entirely within the physical
dimensions are effective spin 2 objects which we consider in this Letter. 
Gravitons with polarizations partially or completely perpendicular to the
physical brane are vector and scalar objects which we will not consider
here since they couple more weakly than the spin 2 type.

The compelling idea that gravity may interact strongly at TeV scale
energies has recently led to a lot of phenomenological activity. TeV
scale gravity can be manifested either directly through real graviton
production, leading to a missing energy signal, or indirectly through
virtual graviton exchanges. Thus, existing and future high energy
colliders can place bounds (or detect) 
on the scale and the number of extra dimensions
in these theories by looking for such 
signals~\cite{hepph9811337},~\cite{wells}~--~\cite{newrizzo}.

Typically, direct signals drop as $(E/M_D)^{\delta+2}$, where $E$ is the
maximum energy carried by the emitted gravitons. Therefore, the best
limits on $M_D$ from the existing experimental data at LEP2, Tevatron and
HERA are obtained for the case $\n=2$.  For example, existing LEP2 data on
$\sigma(e^+e^- \to \gamma + missing~energy)$ already places the bound,
$M_D \gsim 1$ TeV for $\n=2$ via the process $e^+e^- \to \gamma +G$ (see
references~\cite{hepph9811337, wells, hepph9903294}). For $\n=4$ the
limit is $M_D \gsim 700$ GeV. A NLC with c.m. energy $\gsim 1$ TeV can
push this limit up to $M_D \gsim 6$ TeV (for $\n=2$) and $M_D \gsim 4$ TeV
(for $\n=4$) \cite{hepph9811337}. In hadronic colliders, the signal $p \bar
p \to jet + missing~energy$ can proceed by the subprocesses $q \bar q \to
g G$, $q (\bar q) g \to q (\bar q) G$ and $gg \to gG$.  Using these, the
existing Tevatron data on $\sigma(p \bar p \to jet + missing~energy)$
places the limits $M_D \gsim 750$ GeV for $\n=2$ and $M_D \gsim 600$ GeV
for $\n=4$, while the LHC will be able to probe $M_D$ up to $\sim 7$ TeV for
$\n=2$ \cite{hepph9811337,wells}.

The present bounds obtained from indirect signals associated with virtual
graviton exchanges are typically $M_D \gsim 500 -700$ GeV via processes
such as $e^+e^- \to \gamma \gamma,~ZZ,~W^+W^-$ (LEP2) \cite{hepph9902263},
$e^+q \to e^+q$ or $e^+g \to e^+g$ (HERA)
\cite{hepph9812486,hepph9901209}, $p \bar p \to t \bar t + X$ (Tevatron)
\cite{hepph9811501}, and $M_D \gsim 1$ TeV via processes such as $q \bar
q,gg \to \ell^+\ell^-$ (Tevatron) and $e^+e^- \to f \bar f$ (LEP2)
\cite{hepph9811356,hepph9901209}. Future colliders such as the NLC and the
LHC will be able to push these limits to several TeV's through the study
of these signals.  Clearly other new physics can also give rise to similar
signals so while a search for such signals may be used to bound TeV scale
gravitation theories, to clearly identify gravitation as the source
generally requires more extensive analysis, such as the study of angular
distributions distributions of final state particles 
(e.g.~\cite{hepph9811356}).

It should also be noted 
that the predictions in virtual graviton processes 
have some uncertainties since they depend on the sum over the
Kaluza Klein (KK) tower of the massive  excitations which is not
fully determined without knowing the full quantum gravity theory.

In this paper we investigate another possible direct signal of strongly
coupled low energy gravity via the process $e^+e^- \to e^+e^- G$ ($G$=spin
2 graviton) which proceeds predominantly through the t-channel $\gamma
\gamma$ (or $ZZ$) fusion subprocesses $\gamma \gamma (ZZ) \to G$, as well
as the related process $e^-\gamma\to e^- G$. In the case of the two photon
fusion, the photons are produced virtually and since these photons tend to
be collinear, the process $e^+e^- \to e^+e^- G$ is significantly enhanced
compared to $s$-channel processes. We find that the resulting signal is
robust and useful for detecting or constraining some low energy gravity
scenarios at the energy scales of a future NLC. 

Using this method, a $1~TeV$ electron-positron collider with integrated 
luminosity of $200~fb$ is sensitive to a gravitational scale of $2.8~TeV$ 
for 2 extra dimensions and $1.5~TeV$ for 6 extra dimensions. The main 
factors limiting the sensitivity are: 
\begin{enumerate}
\item 
The Standard Model background 
$e^+e^-\to e^+e^-\nu_e\bar\nu_e$ which leads to a final state with the 
same experimental signature 
\item
The necessity to observe the two electrons with a significant $P_T$ in 
order to infer the existence of a missing particle (i.e. the graviton).
\end{enumerate}

The rate of graviton production would be greatly enhanced if, instead of
colliding two virtual photons, one were to collide two real photons
produced via backscattering from the electron
beams~\cite{Ginzburg:1983vm}. Unfortunately, in this case the process
would be $\gamma\gamma\to G$ which would have no signal in the detector. 
On the other hand, if a laser only backscatters from one of the electron
beams, then the process is $e^-\gamma\to e^- G$ and would give a signal of
a single electron with a large unbalanced transverse momentum.

In this case the main limiting factor is the Standard Model background
primarily from $e^-\gamma\to e^-\nu_e\bar\nu_e$.  An election-positron
collider with center of mass energy of $1~TeV$ which is converted to an
electron-photon collider will thus be sensitive to a new gravitational
scale of $10.4~TeV$ in the case of 2 extra dimensions and $2.7~TeV$ in the
case of 6 extra dimensions.

In section 2 we discuss the case of graviton production through 
$\gamma\gamma$ fusion at electron-positron colliders. In section 3 we 
consider the case of $e^-\gamma\to e^-~+~graviton$ at an electron-photon 
collider and in section 4 we give our concluding remarks.

\section{Graviton Production by Two Photon Fusion}

Let us first consider the excitation of spin 2 graviton modes through
photon-photon and $ZZ$ fusion. Such a process could be probed at an
$e^+e^-$ collider where the effective photon luminosity is generated by
collinear photon emission. The complete process is therefore $e^+e^-\to
e^+ e^- G$ through the diagram shown in Fig.~1.  In principle  other
diagrams where the graviton is attached to the fermion lines or directly
to the gauge-fermion vertex will also contribute, but the process in
Fig.~1 should be dominant due to the enhancement of collinear gauge boson
emission.

The cross section
of the photon-photon fusion  process may be estimated through the 
Weiszacker-Williams leading
log approximation~\cite{peskin_schroeder}.  Thus if $\sum|\fm(\hat s)|^2$
is the matrix element for $\gamma\gamma\to G$, where $G$ is a graviton of
mass $m=\sqrt{\hat s}$, then in this approximation the total cross 
section for $e^+e^-\to e^+e^- G$ is given by:

\begin{eqnarray}
\sigma(e^+e^-\to e^+e^-G)
={\pi\eta^2\over 4 s}
\int_0^1 
{f(\omega)\over \omega} 
D(\omega s)
\sum|\fm(\omega s)|^2 d\omega
\label{wwaprox}
\end{eqnarray}

\noindent
where $s$ is the center of mass energy of the collision,
\begin{eqnarray}
f(\omega)=\left [ (2+\omega)^2\log(1/\omega)
-2(1-\omega)(3+\omega)\right ] /\omega 
\nonumber
\end{eqnarray}
\noindent
and
$\eta=\alpha \log\left [s/(4m_e^2)\right ] /(2\pi)$.

Using the effective Lagrangian for the $G\gamma\gamma$ coupling derived
in~\cite{wells,taohan}, we obtain: 

\begin{eqnarray}
\sum|\fm(\hat s)|^2 
=2 {\hat s^2\over \BAR M_P^2}
\label{matel}
\end{eqnarray}

\noindent
Note that the explicit dependence on $\BAR M_P$ will cancel when 
multiplied by the density of graviton states.
This is typical of reactions involving real graviton emission.
We therefore obtain the total cross section in this approximation:

\begin{eqnarray}
\sigma_{\gamma\gamma}(e^+e^-\to e^+e^-G)
=
{\alpha^2\over 16\pi s}
S_{\delta-1} 
\left [{\sqrt s\over M_D}\right ]^{\delta+2}
F_{\delta\over 2} 
\log^2\left [ {s\over 4 m_e^2}\right ]
\end{eqnarray}

\noindent
where $F_k=\int_0^1 f(\omega)\omega^k d\omega$.

In Fig.~2, the solid curves give the total cross section as a function of
$s$ given $M_D=1$ TeV for $e^+e^-\to e^+e^-G$ in the cases where
$\delta=$2 and 6 (corresponding to the upper and lower solid curves) while
the thin dashed curve is the cross section for $\mu^+\mu^-\to \mu^+\mu^-G$
with $\delta=2$ which would be applicable to a muon collider.
We see, for example, that if the gravitational interactions scale is 
1 TeV, hundreds (thousands) 
of such $e^+e^-G$ events will be produced already at 
LEP2 (at a 500 GeV NLC).     

Experimental considerations 
suggest, however, that perhaps the full cross section
which is given in the above is not observable.  Gravitons couple very
weakly to normal matter and thus a radiated graviton will not be detected
in the detector.  Therefore, the signature for the reaction would be

\begin{eqnarray}
e^+e^-\to e^+e^-~+~missing~mass. 
\nonumber
\end{eqnarray}

\noindent Since this cross section is dominated by emission of photons at
a small angle, the outgoing electrons will therefore also be deflected by
a small angle. Although one can expect that the electrons will suffer an
energy loss, a significant portion of the electrons will not be deflected
out of the area of the beam pipe and so may not be directly detected.  To
obtain a more realistic estimate one must therefore select events where
the electron is deflected enough to be detected. Moreover, there is a
Standard Model (SM) background to this signal from the processes
$e^+e^-\to e^+e^-\nu_\ell\bar\nu_\ell$, $\ell = e,~\mu,~\tau$.  We
performed the exact calculation of this background by means of the CompHEP
package \cite{comphep}. This background is found to be dominated by the
$e^+e^-\nu_e \bar\nu_e$ final state; it is $\sim 420~fb$ for
$\sqrt{s}=500$ GeV and $\sim 360~fb$ at $\sqrt{s}=1$ TeV (out of which
$\sim 90\%$ is from $e^+e^-\nu_e \bar\nu_e$), including all neutrino
flavors and when no cuts are imposed.
\footnote{There is, in principle,
additional reducible background, e.g., two photon production when one is
lost down the beam pipe. 
Such a background is tied to the detailed specifications of the detector. 
We therefore consider it
beyond the scope of this paper, however, it should be noted
that it is expected to be much smaller than the SM irreducible background.}
%
%
%
Note that this background includes subprocesses where there are
intermediate states of two gauge bosons such as $e^+e^-\to W^+W^-\to e^+
\nu_e e^- \bar\nu_e$ and $e^+e^-\to Z\gamma\to \nu\bar\nu e^+ e^-$.
Tau-lepton pair production can also lead to a background with four
neutrinos in the final state, specifically, $e^+e^-\to\tau^+\tau^-$ where
each of the $\tau$-leptons decays leptonically to $e\nu\bar \nu$. We find,
however, that this gives a contribution which is about two orders of
magnitude smaller than $ee\nu\nu$. 
%
%

Let us now consider three possible methods for detection of this signal. 
First, one could take advantage of the fact that a significant amount of
energy present in the initial collision is lost to the unobservable
graviton. 
In Fig.~3, the normalized missing mass distribution is shown as
a function of $\omega=\hat s/s$ where $\hat s$ is the missing mass squared
of the graviton.  In this approximation, this distribution is not changed
by the value of $\sqrt{s}$, $M_D$ or any systematic cut imposed on the
transverse momentum $P_T$ of the outgoing electrons.  
The distribution is
shown for $\delta=2$ (solid), $\delta=4$ (dashed), $\delta=6$ (dotted) and
$\delta=8$ (dot-dash), where it is evident that, for the cases 
$\delta=4,~6$ and 8, the missing mass carried by the graviton is 
predominantly concentrated at high $\omega$ values.
In contrast, the 
missing mass spectrum for the background discussed above (in particular, 
for the dominating background process $e^+e^- \to e^+e^- \nu_e \bar\nu_e$) 
is roughly flat with no significant peaks.
Therefore, to obtain a bound on $M_D$ it may also be useful
to consider a cut on $\omega$ for the cases 
$\delta >2$. For example, for $\sqrt s=1$ TeV, the cut $\omega<0.16$
reduces the background 
by a factor of $\sim 0.38$, 
while the signal is reduced by a factor of 0.42 in the
case of $\delta=2$, 0.82 in the case of $\delta=4$, 0.96 in the case of
$\delta=6$ and 0.99 in the case of $\delta=8$.  
We will not consider further such a cut on $\omega$. 
We note however,  
that if a signal is seen, the missing mass distributions in Fig.~3 
may help distinguish these 
theories from other new physics candidates and also help to determine how 
many extra dimensions are present.

In principle it might be possible to separate the
reduced energy electrons from the outgoing electrons of the collision at a
$e^+e^-$ collider through downstream dipole magnets but the large
bremsstrahlung radiation generated by the disruption of the collision
probably makes such an electron difficult or impossible to detect. At a
muon collider, perhaps a Roman Pot could find reduced energy muons which
were deflected from the main beam, however the decay electrons in the muon
collider environment may make this difficult also. Clearly experimental
innovations are required to detect the full cross section and we will not
consider this further.

Secondly, if both of the electrons are given enough of a transverse
momentum that they may be detected in the detector or the end-caps, events
of the desired type may be identified. Using the leading log
approximation, one can use Eqn.~\ref{wwaprox} with $\eta$ replaced by
$\hat \eta(P_{Tmin})=\alpha \log\left [s/(4P^2_{Tmin})\right ] /(2\pi)$
where $P_{Tmin}$ is the minimum transverse momentum of the outgoing
electron which is accepted.  If one imposes this cut on the two outgoing
electrons one obtains the cross section as a function of $\sqrt{s}$ shown in
Fig.~2 with the dotted curve for the case of $P_{Tmin}=10$~GeV with
$M_D=1$ TeV and $\delta=2$, while the heavy dot-dot-dash curve is for
$\delta=4$.  These curves would be the same at both electron and muon
colliders since the transverse momentum cut is well above the lepton mass.
The missing mass spectra under this cut should also correspond to the
curves shown in Fig.~3.

Thirdly, one could identify events where only one of the electrons has a
transverse momentum greater than $P_{Tmin}$.  This would in effect be
replacing $\eta^2$ in Eqn.~\ref{wwaprox} with
$\eta_{eff}^2=\eta(P_{Tmin}) (2\eta-\eta(P_{Tmin}))$. 
The resultant cross sections
are shown in Fig.~2 with the dot-dash curve for $P_{Tmin}=20$~GeV.  In
this case, the energy of the detected electron will be markedly reduced
from the beam energy since the graviton mass distribution increases at
high masses.  In Fig.~4 we show the normalized missing energy ($E_{miss}$)
spectrum as a function of $x=2E_{miss}/\sqrt{s}=E_{miss}/E_{beam}$ for the
detected electron where $\delta=2$ (solid), $\delta=4$ (dashed),
$\delta=6$ (dotted) and $\delta=8$ (dot-dash). In this leading log
approximation, the curves of Fig.~4 are largely independent of $P_{Tmin}$. 
For instance, if we impose the cut $x>0.2$, 
the signal is reduced by a factor of 0.72 in
the case of $\delta=2$, 0.93 in the case of $\delta=4$, 0.99 in the case
of $\delta=6$ and 0.997 in the case of $\delta=8$. 
We will not use this cut in our numerical results below, 
however, again, if a signal is seen, the 
missing energy distributions in Fig.~4 
may provide for an extra handle in resolving the origin 
of such a signal.


Let us now consider the related process $e^+e^- \to ZZ e^+ e^- \to e^+ e^-
G$ which can likewise be estimated by the effective vector boson leading
log approximation.  In general the cross section is given by a sum over
cross sections for $ZZ\to G$ in various helicity combinations together
with the helicity dependent structure functions given in ~\cite{evba}.
Here there is considerable simplification since in this approximation
where the boson momenta are taken collinear with their parent leptons, the
only amplitude which contributes are the cases where the bosons are
transverse and of opposite helicities. As with the photon case, we use the
effective Lagrangian from~\cite{wells} and obtain the cross section in
this approximation:

\begin{eqnarray}
\sigma_{ZZ}(e^+e^-\to e^+e^-G)
=
{y^2\alpha^2\over 16\pi s}
S_{\delta-1} 
\left( {\sqrt s\over M_D} \right )^{\delta+2} 
\left [ F_{\delta\over 2}^Z(s)+z^2 H_{\delta\over 2}^Z(s) \right ]
\log^2 \left( {s\over M_Z^2} \right )
\end{eqnarray}

\noindent
where

\begin{eqnarray}
&&x_w=\sin^2\theta_w,\ \ \ \ 
y={1-4x_w+8x_w^2\over 8 x_w (1-x_w)}, \ \ \ \
z={1-4x_w\over 2(1-4x_w+8x_w^2)},
\nonumber\\
&&F_k^Z(s)=\int_{4 m_Z^2\over s}^1 \omega^k f(\omega) d\omega
\nonumber\\
&&H_k^Z(s)=-\int_{4 m_Z^2\over s}^1 4 \omega^{k}
\left[(4+\omega)\log(\frac{1}{\omega}) - 4(1-\omega) \right]
\end{eqnarray}

\noindent 
and $f(\omega)$ is defined as for the case of photons.

In Fig.~2 the thick dashed curve shows the total cross section for this
process given $M_D=1$ TeV and $\delta=2$. 
Clearly, the $ZZ$-fusion cross section is much smaller than 
the two photon process.\footnote{We note that, for massive vector 
bosons, the effective vector boson approximation in leading log tends to 
overestimate the cross section, in particular, the cross section coming 
from fusion of transversely polarized gauge bosons, see e.g., Johnson 
{\it et al.} in \cite{evba}. However, since the photon-photon process is 
much 
larger than the $ZZ$ even when the latter is calculated in the 
leading log 
approximation, this effect is negligible for our numerical results.}
Moreover, 
this cross section is flat in
$P_T$ for $P_T<O(m_Z)$ and therefore $O(10~{\rm GeV})$ cuts in $P_T$ of the
outgoing leptons will not reduce this greatly.  For similar reasons the
cross section at a $\mu\mu$ collider will be the same.

In Table~1 we consider the limits that may be placed on theories with
extra dimensions using these $e^+e^-\to e^+e^-G$ processes in 
case no such signal is detected. We consider
three possible accelerator scenarios: 
$\sqrt{s}=200$ GeV and a total 
integrated luminosity of $2.5~fb^{-1}$ (for LEP-200);
$\sqrt{s}=500$ GeV and a total integrated luminosity of $50~fb^{-1}$; 
$\sqrt{s}=1$ TeV and a total integrated luminosity of 
$200~fb^{-1}$. These last two cases correspond to a future NLC.
For $\delta=2$, $4$ and $6$ we 
consider detection either via the full cross section (if that were somehow
observable) or via the signal with the cut $P_{Tmin}=10$ GeV on just one
outgoing electron or both outgoing electrons. 

We define the lower limit on $M_D$ in each case to be the value which will
yield a signal with a statistical significance 
of $3\sigma$ above the background by requiring:

\begin{eqnarray}
\frac{\sigma^T - \sigma^{SM}}
{\sqrt{\sigma^T}} \times L >3 \label{limit} ~,
\end{eqnarray}

\noindent where $\sigma^T$ is the total cross section for $e^+e^- + missing
 ~ energy$ production and $\sigma^{SM}$ is the SM contribution to this signal.
$L$ is the luminosity of the collider and we also require 
(for the given $L$) at least 
10 such $e^+e^-G$ 
events above the SM background for the given lower 
bound on $M_D$.\footnote{In fact, we find that, using the $3\sigma$ 
lower bounds on $M_D$ as given in Table 1 and the given colliders 
luminosities, about $50-100$ $e^+e^- G$ events 
will be produced at LEP2 energies, while hundreds - thousands such events 
will emerge at 500 and 1000 GeV NLC, 
for all three values of $\delta$ considered in Table 1.}

As can be seen, in the case of two extra dimensions and using the signal
with the two electron $P_{Tmin}$ cuts, a limit of about 600 GeV may be
placed on $M_D$ at the $200$ GeV collider; using the $500$ GeV collider a
limit of about 1.7 TeV may be obtained and with a $1$ TeV collider a limit
of about 2.8 may result. Thus, in general, a given collider (out of the
three scenarios above) can place a $3\sigma$ bound on $M_D$ of about three
times its c.m. energy.  Obviously, with less stringent cuts and/or using a
single high $P_T$ lepton tag, the lower limit on $M_D$ may be increased. 
Also, we note that, as expected, the limit on $M_D$ decreases
somewhat as $\delta$ increases and that, as mentioned before, a lower cut
on the missing mass may be of some advantage if $\delta >2$.

\section{Graviton Production by Electron-Photon Collisions}

It has been suggested~\cite{Ginzburg:1983vm} that an electron-positron
collider might be converted to an electron-photon or photon-photon
collider by scattering a laser beam from one or both of the electron
beams. This would produce a great enhancement over the virtual photon
luminosities considered above. If a photon-photon collider were used,
however, the production of gravitons could not be considered as above
since the method requires observation of outgoing electrons to infer that
a graviton was produced. On the other hand, at an electron-photon
collider, the process $e^-\gamma \to e^- G$ would lead to a high
transverse momentum electron in the final state which could be detected.
The signature for such an event would therefore be $e^-\gamma\to e^- +
P_T^{miss}$ where the missing transverse momentum is the same in the lab
and in the $e\gamma$ rest frame.

As shown in Fig.~5, this process could proceed through several diagrams in
addition to the one analogous to the fusion process considered above. This
leads to the differential hard cross section in the $e^-\gamma$ center of
mass frame for producing a graviton of mass $m$: 

\begin{eqnarray}
{d\sigma(m)\over dz}&=&
{
\pi\alpha G_N 
\over
4(1-z^2)
}
\left ( 4(1+z)+x(1-z) \right)
\nonumber\\
&&\times \left ( 5-6x+5x^2+2z(1-x^2)+z^2(1-x)^2 \right )
\label{differential1}
\end{eqnarray}

\noindent
where $z=\cos\theta_{ee}$, $\theta_{ee}$ being the angle between the
initial and final electron momenta in the center of mass frame and
$x=m^2/s$ where $s$ is the center of mass energy of the collision
and $G_N$ is the Newtonian gravitational constant.
This formula is related by crossing 
symmetry
to that derived in~\cite{hepph9811337}
for $e^+ e^-\to \gamma G$. 
This distribution must be convoluted with the density of states
in Eqn.~(\ref{density_of_states}). 

In the experimental setting where the photon beam is produced by laser
backscatter, the energy of the photon in a specific event is not known. 
Since the missing mass is also unknown, the event cannot be fully
reconstructed and the distribution in $z$ cannot be directly observed. It
is more useful therefore to consider the distribution in the transverse
momentum of the electron, $P_T$, or equivalently the missing transverse
momentum $P_T^{miss}$ which is given by 
$P_T=(\sqrt{s}/ 2)(1-x)\sqrt{1-z^2}$.

Convoluting the differential cross section in Eqn.~(\ref{differential1})
with the density of states, we can obtain the following differential cross
section summed over graviton states:

\begin{eqnarray}
{d\sigma\over dz }
=
{\alpha S_{\delta-1}\over 64 s}
\left ({\sqrt{s}\over M_D} \right )^{\delta+2}
{
A_\delta+B_\delta z
+C_\delta z^2+D_\delta z^3
\over 1-z^2}
\end{eqnarray}

\noindent 
where
$A_\delta$, $B_\delta$, $C_\delta$ and $D_\delta$ are given by:

\begin{eqnarray}
A_\delta&=&\int_0^1 (x+4)(5-6x+5x^2)         x^{{\delta\over 2}-1}dx~
\nonumber\\
B_\delta&=&\int_0^1 (28-27x+18x^2-7x^3)      x^{{\delta\over 2}-1}dx
\nonumber\\
C_\delta&=&\int_0^1 3(1-x)(4+x-x^2)          x^{{\delta\over 2}-1}dx
\nonumber\\
D_\delta&=&\int_0^1 (4-x)(1-x)^2             x^{{\delta\over 2}-1}dx
\end{eqnarray}

We must now convolute this distribution with the energy spectrum of the
photons in the collider to obtain the cross section relative to the
$e^+e^-$ luminosity. The distribution 
in terms of the energy fraction $u=E_\gamma/E_e$
given in~\cite{Ginzburg:1983vm} 
for a laser photons scattered from an unpolarized electron beam 
is

\begin{eqnarray}
f(u)=
{
 1/(1-u)+(1-u)-4r(1-r)
\over
(1-4/X-8/X^2)\log(1+X)+1/2+8/X-1/(2(1+X)^2) 
}
\end{eqnarray}

\noindent
where 
$u\leq u_{max}=X/(X+1)$, $r=u/(X(1-u))$ and
$X=4 E_e \omega_0/m_e^2$, $E_e$ being the energy of the scattering
electron beam and $\omega_0$ being the energy of the laser photons.

The total cross section with respect to the $e^-e^+$ luminosity is 

\begin{eqnarray}
\sigma_0=\int_0^{u_{max}} 
\sigma_{e\gamma}(s_0 u) f(u) du
\end{eqnarray}

\noindent
where $s_0$ is the center of mass energy that the $e^+e^-$ system has
without laser scattering.

The number of events is thus $N=\sigma_0 {\cal L}_{ee}$ 
where ${\cal L}_{ee}$
is the integrated luminosity for $e^+e^-$ collisions if the scattering
laser were absent. 

The above result assumes that $\omega_0$ is not so large as to cause
$e^+e^-$ pairs to be created in the scattering. This is equivalent to 
$X\leq 2(1+\sqrt{2})$. We therefore take $X=2(1+\sqrt{2})$ which gives the
hardest spectrum without pair generation. For instance, in the case where
$\sqrt{s_0}=1~TeV$, this value of $X$ corresponds to $\omega_0=0.63~eV$.

In Fig.~6 we show $d\sigma_0/d P_T$ for a collider where 
$\sqrt{s_0}=1~TeV$ in the cases $\delta=2$ (upper solid), $4$ (dotted), 
$6$ (dashed) and $8$ (dot dashed) together with the SM
background (lower solid) calculated with the CompHEP 
package~\cite{comphep}.
In this case the background comes from $e^-\gamma\to e^- \nu_\ell \bar
\nu_\ell$ where $\ell=e$, $\mu$ and $\tau$. As before, the dominant
background is generated by $\ell=e$. 

In Table 2 we give the 3-$\sigma$ limit that may be obtained on $M_D$ for
$\delta=2$, $4$ and $6$ for collider scenarios with $\sqrt{s_0}=500~GeV$
and ${\cal L}=50~fb^{-1}$;  $\sqrt{s_0}=1000~GeV$ and ${\cal
L}=200~fb^{-1}$ and $\sqrt{s_0}=1500~GeV$ and ${\cal L}=200~fb^{-1}$.
We also impose an acceptance cut of $P_T>\sqrt{s_0}/10$.

In the case of $\delta=2$ fairly stringent bounds of 
$5.5~TeV$, 
$10.4~TeV$
and 
$14.0~TeV$ 
can be placed on the scale of gravitational interactions 
for $\sqrt{s_0}=500~GeV$, $1000~GeV$ and $1500~GeV$,
respectively.
This, however falls short of bounds in the $\delta=2$ that
have been obtained from supernova cooling~\cite{astro_star} ($\sim
13~TeV$) and the absence of diffuse cosmic gamma ray
backgrounds~\cite{astro_univ} ($\sim 100~TeV$). Indeed, in a collider with
$\sqrt{s_0}=2~TeV$ and ${\cal L}=200~fb^{-1}$ the corresponding bound is 
$18~TeV$ comparable to the supernova result. For larger $\delta$ where the
astrophysical results do not apply, fairly stringent bounds may be placed
on $M_D$. For instance in the case of $\delta=4$ we obtain the bound of
$2.2~TeV$, 
$4.2~TeV$, 
$5.8~TeV$ 
for $\sqrt{s_0}=500~GeV$, $1000~GeV$ and
$1500~GeV$, respectively.

\section{Conclusion}

In conclusion, we have shown that effective photon-photon and real
photon-electron collisions can produce massive gravitons at appreciable
rates in theories with large extra dimensions where the gravitation scale
is $1-10~TeV$. In the case of two photon fusion, the Standard Model
background from $e^+e^-\to e^+e^-\nu_e\bar\nu_e$ limits the effectiveness
somewhat but, for instance at a $1~TeV$ collider with integrated
luminosity $200~fb^{-1}$ we are sensitive to a gravitation scale of
$M_D=2.8~TeV$ in the case where $\delta=2$ and $1.5~TeV$ in the case where
$\delta=6$ taking a cut of $P_{Tmin}=10~GeV$ on both the outgoing
electrons. 

The case of electron-photon fusion at an electron-photon collider is
limited to a lesser extent by the Standard Model background $e^-\gamma\to
e^-\nu_e\bar \nu_e$. Using this case at an electron-photon collider based
on an electron-positron collider with center of mass energy of $1~TeV$, a
gravitation scale up to $10.4~TeV$ may be probed in the case of $\delta=2$
and $2.7~TeV$ in the case of $\delta=6$.

\section{Acknowledgments}

We are grateful to Jose Wudka for discussions.  One of us (DA) thanks the
UCR Theory Group and the Fermilab Theory Group for hospitality. This
research was supported in part by US DOE Contract Nos.  DE-FG02-94ER40817
(ISU), DE-FG03-94ER40837 (UCR) and DE-AC02-98CH10886 (BNL)

\newpage

\begin{center}
{\Large\bf Table 1}
\end{center}
$$
\begin{tabular}{||c|c|c|c|c||} 
\hline
\multicolumn{5}{||c||}{$\delta=2$}\\ \hline & 
&
{No cut}&
{$P_{Tmin}=10$~GeV}&
{$P_{Tmin}=10$~GeV}\\ 
$\sqrt{s}$ & 
$\int {\cal L} dt$&
{\ }&
{(one electron)}&
{(two electrons)}\\ 
\hline
200~GeV&
2.5~$fb^{-1}$&
1.3~TeV&
1.0~TeV&
0.6~TeV
\\ \hline
500~GeV&
50~$fb^{-1}$&
2.8~TeV&
2.4~TeV&
1.7~TeV
\\ \hline
1000~GeV&
200~$fb^{-1}$&
4.1~TeV&
3.6~TeV&
2.8~TeV
\\ 
\hline
\hline
\multicolumn{5}{||c||}{$\delta=4$}\\ \hline
& 
&
{No Cut}&
{$P_{Tmin}=10$~GeV}&
{$P_{Tmin}=10$~GeV}\\ 
$\sqrt{s}$ & 
$\int {\cal L} dt$&
{\ }&
{(one electron)}&
{(two electrons)}\\ 
\hline
200~GeV&
2.5~$fb^{-1}$&
0.7~TeV&
0.5~TeV&
0.4~TeV
\\ \hline
500~GeV&
50~$fb^{-1}$&
1.6~TeV&
1.4~TeV&
1.0~TeV
\\ \hline
1000~GeV&
200~$fb^{-1}$&
2.5~TeV&
2.3~TeV&
1.9~TeV
\\ 
\hline
\hline
\multicolumn{5}{||c||}{$\delta=6$}\\ \hline
& 
&
{No Cut}&
{$P_{Tmin}=10$~GeV}&
{$P_{Tmin}=10$~GeV}\\ 
$\sqrt{s}$ & 
$\int {\cal L} dt$&
{\ }&
{(one electron)}&
{(two electrons)}\\ 
\hline
200~GeV&
2.5~$fb^{-1}$&
0.5~TeV&
0.4~TeV&
0.3~TeV
\\ \hline
500~GeV&
50~$fb^{-1}$&
1.1~TeV&
1.0~TeV&
0.8~TeV
\\ \hline
1000~GeV&
200~$fb^{-1}$&
1.9~TeV&
1.8~TeV&
1.5~TeV
\\ 
\hline
\end{tabular}
$$

\bigskip
\bigskip

{\bf Table 1:} The $3\sigma$ limits 
on the parameter $M_D$, as defined in Eqn.~\ref{limit},
 are given for $\delta=2$,
4 and 6. In each case three accelerator scenarios are considered:
$\sqrt{s}=200$ GeV, $500$ GeV and $1000$ GeV with luminosities
$2.5~fb^{-1}$, $50~fb^{-1}$ and $200~fb^{-1}$, respectively. 
The signals
considered are based on the total cross section, the cross section with
one electron passing the $P_{Tmin}=10$ GeV cut and the cross section with
both electrons passing the $P_{Tmin}=10$ GeV cut.

\newpage

\begin{center}
{\Large\bf Table 2}
\end{center}
$$
\begin{tabular}{||c|c|c|c|c||} 
\hline
$\sqrt{s_0}$ & 
$\int {\cal L}_{ee} dt$&
$\delta=2$ &
$\delta=4$ &
$\delta=6$ \\ 
\hline
500~GeV&
50~$fb^{-1}$&
5.5~TeV&
2.2~TeV&
1.3~TeV
\\ \hline
1000~GeV&
200~$fb^{-1}$&
10.4~TeV&
4.2~TeV&
2.7~TeV
\\ \hline
1500~GeV&
200~$fb^{-1}$&
14.0~TeV&
5.8~TeV&
3.8~TeV
\\ 
\hline
\hline
\end{tabular}
$$

\bigskip
\bigskip

{\bf Table 2:} 
The $3\sigma$ limits 
on the parameter $M_D$, as defined in Eqn.~\ref{limit},
are given for $\delta=2$,
4 and 6 using the process $e^-\gamma\to e^-G$ where the electron beams are
assumed to be unpolarized.
In each case three accelerator scenarios are considered:
$\sqrt{s_0}=500$ GeV, $1000$ GeV and $1500$ GeV with 
$e^+e^-$
luminosities
$50~fb^{-1}$, $200~fb^{-1}$ and $200~fb^{-1}$, respectively. 
In all cases we apply the cut $P_{Tmin} =\sqrt{s_0}/10$.

\newpage
\begin{center}
{\Large\bf Figure Captions}
\end{center}

\bigskip

\noindent {\bf Figure 1:} The dominant Feynman diagram for $e^+e^-\to e^+
e^-G$ through an effective photon or $Z$ sub-process.

\bigskip

\noindent {\bf Figure 2:} 
The cross sections for $\ell^+ \ell^- \to \ell^+ \ell^- G$
for various values of $\delta$ are shown as a function of
$\sqrt{s}$. 
The solid lines are the total cross sections for 
$e^+ e^- \to e^+ e^- G$ for $\delta=2$ (upper curve)
and $\delta=6$ (lower curve). 
The dotted line is for the case that both the outgoing electrons are subject 
to the cut $P_{Tmin}=10$ GeV and for $\delta=2$. 
The dot-dash line 
is obtained again with
$\delta=2$ but now only  one of the outgoing electrons is subject to the cut
$P_{Tmin}=20$ GeV.
The thick dot-dot-dash line 
is for
$\delta=4$ where both of the outgoing electrons are subject to the
cut $P_{Tmin}=10$ GeV.
The thick dashed line shows the total cross section 
for $\delta=2$ via the $ZZ$ process. 
The dashed line gives the cross section for $\mu^+\mu^-\to\mu^+\mu^-G$
for $\delta=2$ via the $\gamma\gamma$ process. 
In all cases we take $M_D=1$ TeV.

\bigskip

\noindent {\bf Figure 3:} The normalized differential cross section as a
function of the scaled 
missing (graviton) 
invariant mass squared ($\omega=\hat s/s$) for
$\delta=2$ (solid line), $\delta=4$ (dashed line),
$\delta=6$ (dotted line), $\delta=8$ (dot-dash line).  These curves are
not greatly effected by the $P_{Tmin}$ cut, $M_D$ or $s$.

\bigskip

\noindent {\bf Figure 4:} The normalized differential cross section as a
function of the missing energy of the single detected electron.
See also caption to Fig.~3.

\bigskip

\noindent {\bf Figure 5:} The dominant Feynman diagrams 
for $e^-\gamma\to e^- G$.

\bigskip

\noindent {\bf Figure 6:} 
The distribution $d\sigma_0/d P_T$ for the signal and SM
background when $\sqrt{s_0}=1~TeV$ and $M_D=1~TeV$. The photons are
produced by the backscatter of a laser where $X=2(1+\sqrt{2})$ and the
electron beams are taken to be unpolarized. The signal is shown for
$\delta=2$ (upper solid curve); 
$\delta=4$ (dotted curve);
$\delta=6$ (dashed curve) and
$\delta=8$ (dash dot curve). The SM background from 
$e^-\gamma\to e^-\nu_\ell\bar\nu_\ell$ is shown with the lower solid
curve.

\newpage

\
\begin{figure}
\vspace*{0 in}
\hspace*{+1.0 in}
\epsfxsize 3.0 in
\mbox{\epsfbox{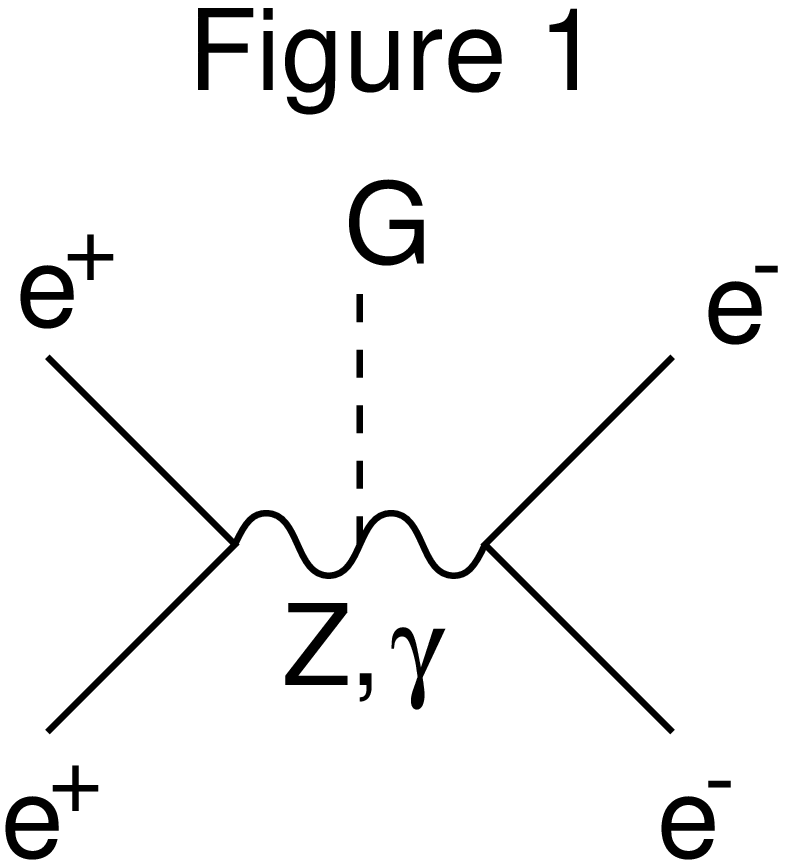}}
\end{figure}

\newpage

\
\begin{figure}
\vspace*{0 in}
\hspace*{-1.0 in}
\epsfxsize 7.0 in
\mbox{\epsfbox{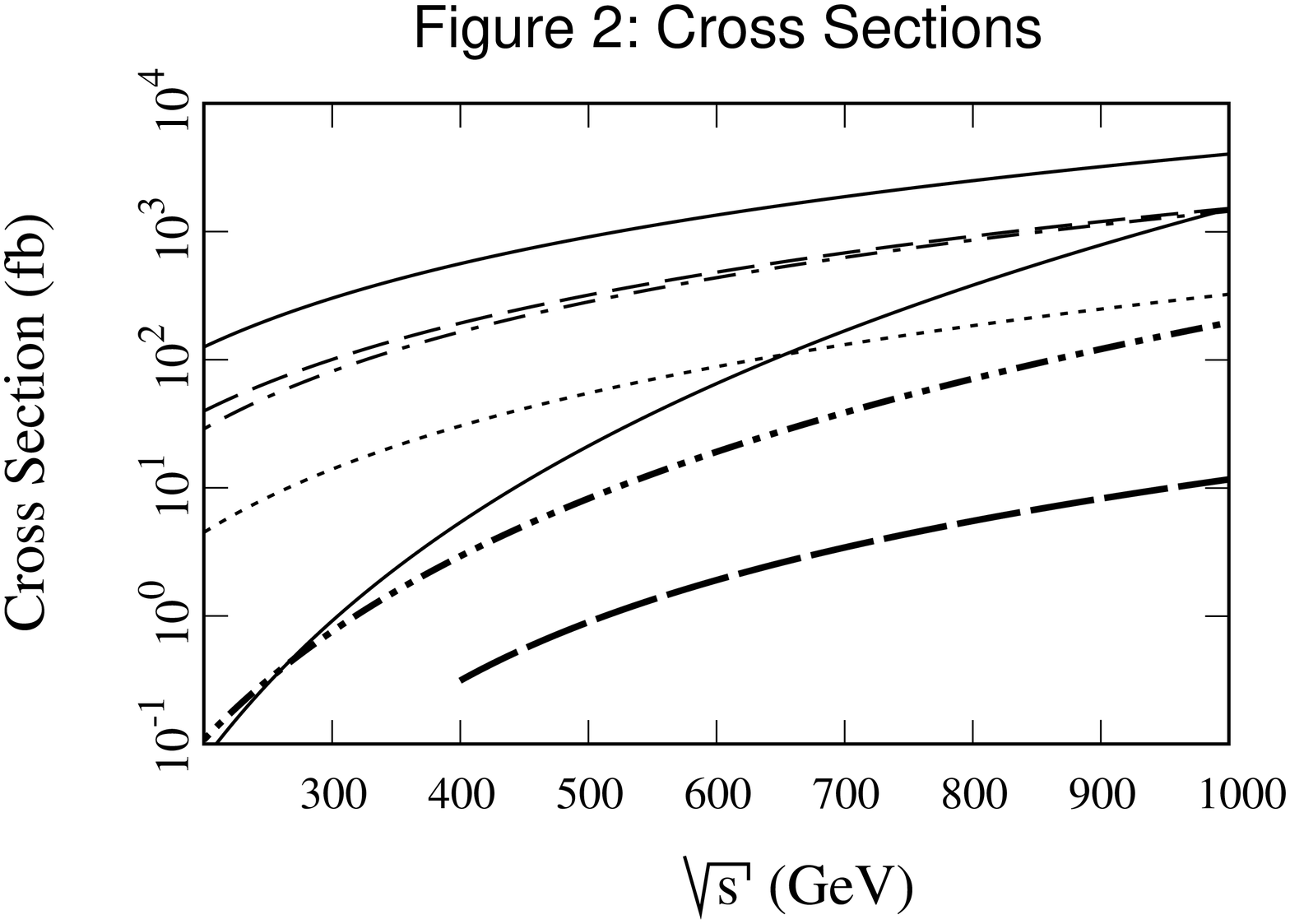}}
\end{figure}

\newpage

\
\begin{figure}
\vspace*{0 in}
\hspace*{-0.5 in}
\epsfxsize 6.0 in
\mbox{\epsfbox{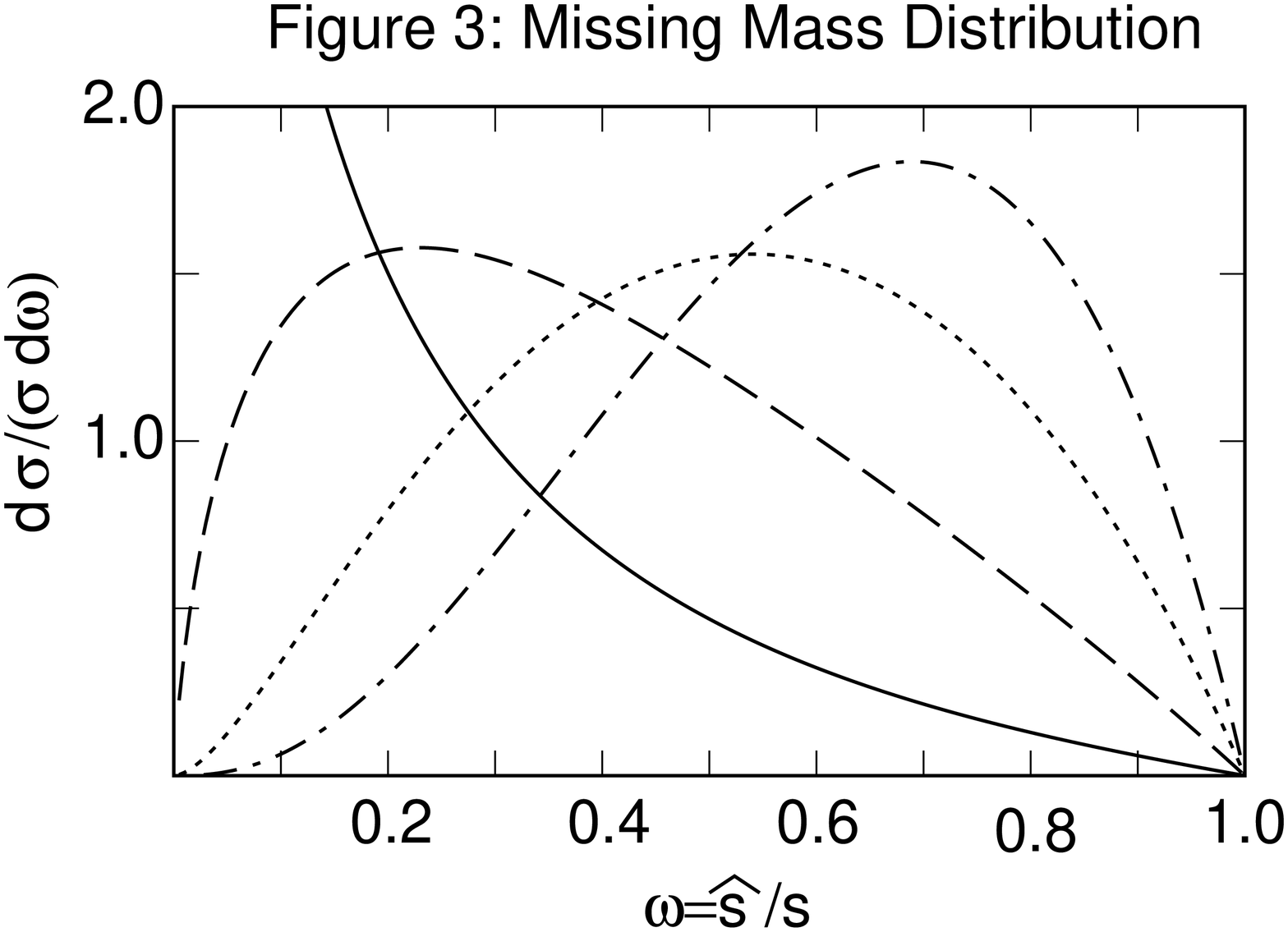}}
\end{figure}

\newpage

\
\begin{figure}
\vspace*{0 in}
\hspace*{-0.5 in}
\epsfxsize 6.0 in
\mbox{\epsfbox{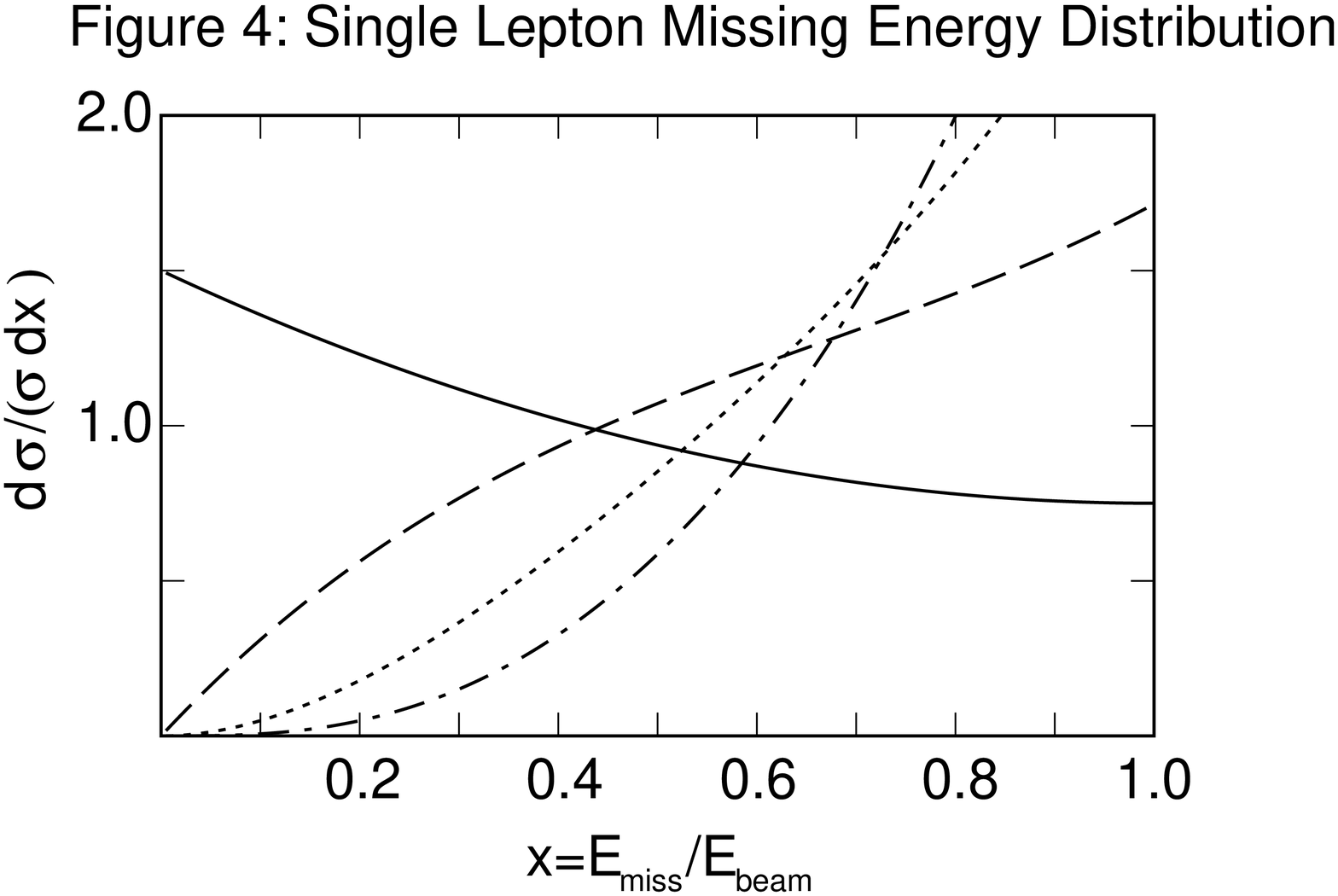}}
\end{figure}

\
\begin{figure}
\vspace*{0 in}
\hspace*{-0.5 in}
\epsfxsize 4.5 in
\mbox{\epsfbox{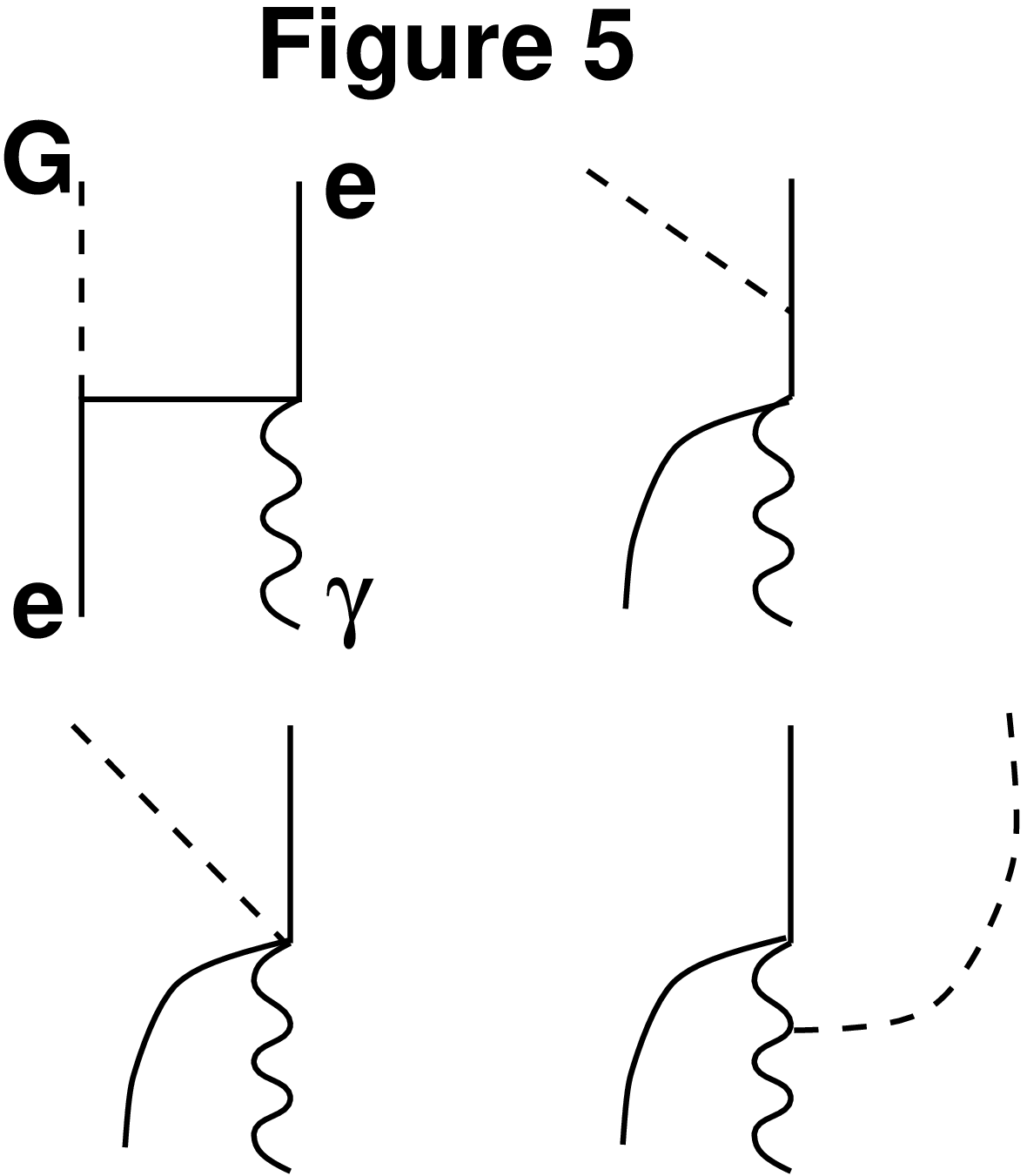}}
\end{figure}

\newpage

\
\begin{figure}
\vspace*{0 in}
\hspace*{-0.5 in}
\epsfxsize 6.5 in
\mbox{\epsfbox{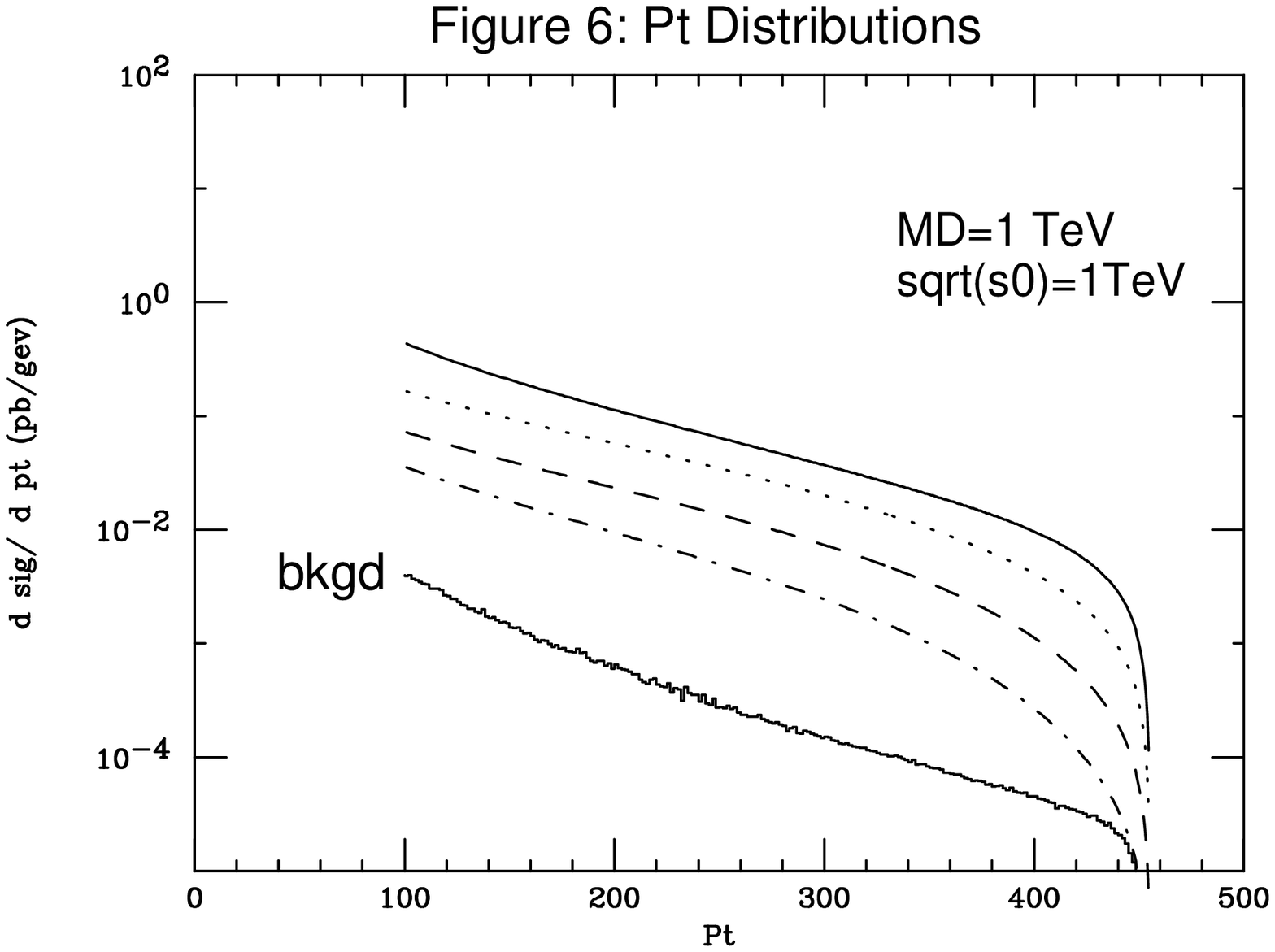}}
\end{figure}


\newpage






\begin{thebibliography}{99}





\bibitem{mtheory}
P.~Horava and E.~Witten, Nucl. Phys. {\bf B460}, 506 (1996);
{\it ibid.} Nucl. Phys. {\bf B475}, 94 (1996);
E.~Witten, Nucl. Phys. {\bf 471}, 135 (1996);
I.~Antoniadis, Phys. Lett. {\bf B246}, 377 (1990);
P.~Ginsparg, Phys. Lett. {\bf B197}, 139 (1987).




\bibitem{add1}
N.~Arkani-Hamed, S.~Dimopoulos and G.~Dvali,
Phys. Lett. {\bf B429}, 263 (1998);
I.~Antoniadis, N.~Arkani-Hamed, S.~Dimopoulos and G.~Dvali,
Phys. Lett. {\bf B436}, 257 (1998).




\bibitem{add2} J.~Lykken, Phys. Rev. {\bf D54}, 3693 (1996); J. Dienes, E.
Dudas and T. Ghergetta, Phys. Lett. {\bf B436}, 55 (1998). G.~Shiu and
S.H.~Tye, Phys. Rev. {\bf D58}, 106007 (1998);
N.~Arkani-Hamed, S.~Dimopoulos and J.~March-Russell, hep-th/9908146, 
hep-th/9809124.




\bibitem{hepph9811337}
E.A.~Mirabelli, M.~Perelstein and M.E.~Peskin,
Phys. Rev. Lett. {\bf 82}, 2236 (1999).


\bibitem{cavin}
V.~P.~Mitrofanov and O.~I.~Ponomareva, Sov. Phys. JETP {\bf 67}, 1963 (1988);
J.~C.~Long, H.~W.~Chan and J.~C.~Price, 
Nucl. Phys. {\bf B539}, 23 (1999) and references therein.




\bibitem{wells}
G.F.~Giudice, R.~Rattazzi and J.D.~Wells,
Nucl. Phys. {\bf B544}, 3 (1999).


\bibitem{taohan}
T.~Han, J.D.~Lykken and R.~Zhang,
hep-ph/9811350 (1998).



\bibitem{convention} In \cite{taohan} the scale of the extra dimension is
parameterized by $M_S$ which is related to $M_D$ via
$M_D^{\delta+2}=(S_{\delta-1}/16\pi)M_S$. 




\bibitem{prd59_105002} 
S.~Nussinov and R.~Shrock, Phys. Rev. {\bf D59}, 105002 (1999).



\bibitem{hepph9811356}
J.L.~Hewett,
hep-ph/9811356 (1998).




\bibitem{Ginzburg:1983vm}
I.F.~Ginzburg, G.L.~Kotkin, V.G.~Serbo and V.I.~Telnov,
Nucl. Instr. Meth. {\bf 205}, 47 (1983);
I.F.~Ginzburg, G.L.~Kotkin, S.L.~Panfil, V.G.~Serbo and V.I.~Telnov,
Nucl. Instr. Meth. {\bf A219}, 5 (1984).






\bibitem{hepph9811501}
P.~Mathews, S.~Raychaudhuri and K.~Sridhar,
hep-ph/9811501 (1998).


\bibitem{hepph9812486}
P.~Mathews, S.~Raychaudhuri and K.~Sridhar,
hep-ph/9812486 (1998).



\bibitem{hepph9901209}
T.G.~Rizzo,
hep-ph/9901209 (1999).



\bibitem{hepph9902263}
K.~Agashe and N.G.~Deshpande,
hep-ph/9902263 (1999).



\bibitem{hepph9903294}
K.~Cheung and W.~Keung,
hep-ph/9903294 (1999).


\bibitem{newrizzo}
T.~G.~Rizzo, hep-ph/9903475 (1999).




\bibitem{peskin_schroeder} {\it See e.g.}, M.~Peskin and D.~Schroeder. ``An
Introduction to Quantum Field Theory'', p.~578, Addison-Wesley Publishing
Company (1995). 

\bibitem{comphep} 
P. A. Baikov {\it et al.}, Physical Results by means of CompHEP, 
in Proceedings of X Workshop on High Energy Physics and Quantum Field
Theory (QFTHEP-95), eds. B. Levtchenko and V. Savrin, Moscow, 1996, p.101, 
hep-ph/9701412; E. E. Boos {\it et al.}, hep-ph/9503280. 
The CompHEP package Version 33 was downlowded from the Web page: 
http://theory.npi.msu.su/comphep.html. 

\bibitem{evba} {\it See e.g.}, S. Dawson, Nucl.\ Phys.\ {\bf B249}, 42 (1985);
P.~W.~Johnson, F.~I.~Olness and W.~K.~Tung, Phys.\ Rev.\ {\bf D36}, 291
(1987); R.P. Kauffman, Phys.\ Rev.\ {\bf D41}, 3343 (1990);  V.~D.~Barger
and R.~J.~N.~Phillips, ``Collider Physics'', Addison-Wesley publishing
company (1987). 

\bibitem{astro_star}
S.~Cullen and M.~Perelstein,
hep-ph/9903422;
V.~Barger, T.~Han, C.~Kao and R.J.~Zhang,
hep-ph/9905474.



\bibitem{astro_univ}
L.J.~Hall and D.~Smith,
hep-ph/9904267.


\end{thebibliography}
\end{document}